\documentclass[runningheads]{llncs}
\usepackage{graphicx}
\usepackage{amsmath}
\usepackage{amsfonts}
\begin{document}
\title{Winning the Ransomware Lottery
\subtitle{A Game-Theoretic Approach to Preventing Ransomware Attacks}
\thanks{Funded in part by the Auerbach Berger Chair in Cybersecurity held by Spiros Mancoridis, at Drexel University}
}
\titlerunning{Winning the Ransomware Lottery}
\author{Erick Galinkin\inst{1, 2}\orcidID{0000-0003-1268-9258}}
\authorrunning{E. Galinkin}

\institute{Rapid7, Boston MA 02114, USA \\ \and 
Drexel University, Philadelphia PA 19104, USA\\
\email{erick\_galinkin@rapid7.com}}

\maketitle             
\begin{abstract}
Ransomware is a growing threat to individuals and enterprises alike, constituting a major factor in cyber insurance and in the security planning of every organization.
Although the game theoretic lens often frames the game as a competition between equals -- a profit maximizing attacker and a loss minimizing defender -- the reality of many situations is that ransomware organizations are not playing a non-cooperative game, they are playing a lottery.
The wanton behavior of attackers creates a situation where many victims are hit more than once by ransomware operators, sometimes even by the same group.
If defenders wish to combat malware, they must then seek to remove the incentives of it.
In this work, we construct an expected value model based on data from actual ransomware attacks and identify three variables: the value of payments, the cost of an attack, and the probability of payment.
Using this model, we consider the potential to manipulate these variables to reduce the profit motive associated with ransomware attack.
Based on the model, we present mitigations to encourage an environment that is hostile to ransomware operators.
In particular, we find that off-site backups and government incentives for their adoption are the most fruitful avenue for combating ransomware.

\keywords{Security \and Malware \and Economics \and Ransomware \and Incentives \and Backups}
\end{abstract}
\section{Introduction}
Ransomware is a family of malware that encrypts files on a system and demands payment for the ability to decrypt these files.
Although proof of concept ransomware has existed since at least 1996~\cite{young1996cryptovirology}, modern ransomware tactics result from CryptoLocker's revolutionary use of Bitcoin for payment~\cite{liao2016behind}.
This innovation has allowed ransomware actors to perpetrate increasingly sophisticated attacks, including the 2017 WannaCry attack~\cite{mohurle2017brief} -- an attack whose effects, according to ransomware payment tracker Ransomwhere\footnote{https://ransomwhe.re} are still being felt today.
We have seen a pivot in targeting, from the wanton use of exploit kits and watering hole attacks that largely affected end users to the current increase in enterprise victims~\cite{o2019symantec} by way of malicious loaders and initial access brokers~\cite{larson2021first}.

The threat of ransomware grows larger year after year, with a spate of recent attacks including on the Colonial pipeline~\cite{morrison2021How} and the Kaseya supply chain attack~\cite{ap2021Scale} demonstrating the devastation and real-world impact of the issues. 
The Ransomware Task Force report~\cite{rtf2021combating} identifies the goal of disrupting the ransomware business model as an important goal.
This goal is uniquely important, since ransomware is so often an attack of opportunity -- akin to a mugging or kidnapping -- and not the sort of highly-targeted attack that is often expected from sophisticated adversaries.
We frame the problem in a new way, as the attacker is not playing a single game against a single defender.
Rather, attackers seek to find vulnerable victims wherever they may be, and so instead of playing a game with attackers, we view the problem from the attacker point of view.
To this end, we suggest that defenders should consider the problem of ransomware and ransomware payments in particular as analogous to an attacker playing a lottery instead of a strategic game between equals.

\section{Related Work}
In recent years, considerable research has been done on the game theory of ransomware payments.
The earliest relevant work on the topic appears to be by Spyridopoulos \textit{et al.}~\cite{spyridopoulos2015game}, who found a Nash equilibrium balancing potential costs of mitigation with the cost of a successful attack. 
Leveraging epidemiologically-inspired models of malware spread, this work considered the equilibria of available defender strategies.
The game is constructed under a unified proliferation model, with infection, immunization, and disinfection rates that informed the strategies of the players.
These player's payoffs were then computed for a set of strategies given the parameters controlled by the attacker and the defender -- the infection rate, patch rate, removal rate, and the rate of both patching and removal.
Spryidopoulos \textit{et al.}'s work informed defenders how to approach ransomware worm attacks and defined the optimal strategy for the defender.

The work of Laszka \textit{et al.}~\cite{laszka2017economics} was the first to consider the economics of ransomware using models that reflect the similarity of ransomware to kidnapping and ransom.
They developed an economic model of the interaction between attackers and victim organizations, and studied that model to minimize the economic impact to those organizations.
Primarily, the work focused on the cost-benefit of investing in backup solutions, a recommendation that is still widely regarded as the best way to prepare for ransomware attacks~\cite{rtf2021combating}.
Laszka \textit{et al.} also showed how coordinated backup investments can deter ransomware attackers in particular -- a novel insight in the literature.
Our work borrows from their recommendations and builds on this existing literature, but we differ in our approach to the game-theoretic model.

Caporusso \textit{et al.}~\cite{caporusso2018game} also built upon the kidnap and ransom literature, leveraging a negotiation model represented as an extensive-form game.
This work dealt with ransomware in cases where renegotiation of the ransom is possible, a surprisingly common phenomenon that has been seen with some ransomware operators~\cite{monroe2021how} -- though other ransomware operators refuse to negotiate.
Caporusso \textit{et al.} identified the post-attack dynamics between the human victim and the human ransomware operator, acknowledging that there are substantial human factors outside of ransom negotiation to be made in the decision making process. 

Cartwright \textit{et al.}~\cite{cartwright2019pay} grappled with the question of whether or not to pay a ransom at all.
Their work largely built upon the earlier paper of Laszka \textit{et al.} and framed the problem of ransomware under the lens of kidnap and ransom. 
It did so by building upon two existing kidnapping models, those of Selten~\cite{selten2013models}, and Lapan and Sandler~\cite{lapan1988bargain}.
The Selten model informed the optimal ransom to be set by the attacker, while the model of Lapan and Sandler aided in deciding whether or not victims should take action to deter the kidnapping in the first place.
In contrast to this work, we present a novel approach to the game and develop a model under a differing set of assumptions.

\section{Probability and Lotteries} \label{sec:prob}
In common parlance, ``lottery'' typically refers to a form of gambling where a player purchases a ticket at some nominal cost with a fixed set of different numbers. 
Then, another set of numbers with the same size is drawn at random without replacement.
After this draw, some reward that confers some amount of utility may be given depending on how many numbers in the randomly drawn set match the set on the purchased ticket.

Mathematically, we can formalize a lottery as follows:
Let $X$ be a set of prizes, $X = \{x_1, ..., x_n\}$, that confers some utility.
From this set of prizes, we define a lottery $L = \{p_i, ..., p_n\}$ over the set of prizes such that for each $x_i \in X$, there is a corresponding $p_i \geq 0$, and $\sum_{i = 1}^n p_i = 1$.
There is also some cost $c \geq 0$ to enter the lottery.
Then, for each of the prizes, there is some utility $u(x_i)$ that the agent derives from receiving that prize, and their expected utility over the lottery is then $\sum_{i=1}^n p_i u(x_i) - c$.
In the ransomware context, a prize $x$ corresponds to a payment to a ransomware operator, and $p$ is the probability that a victim will pay that amount.

The optimal ransom value for $x$ has been explored in other work~\cite{cartwright2019pay} so we instead deal with the binary probability that a victim will pay or not pay, assuming that the optimal ransom value is set.
In our ransomware lottery, we thus define 2 probabilities: $p_{\text{win}}$, when a victim pays a ransom and $p_{\text{lose}} = 1 - p_{\text{win}}$, when a victim does not.
For simplicity in this initial model, we incorporate the probability that the attack is not successful into $p_{\text{lose}}$.
There is, as mentioned, also some small cost $c$ associated with launching the ransomware attack.

Conveniently for ransomware operators, $c$ is quite small, and $x_{\text{win}}$ can be quite large, as we discuss in Section~\ref{sec:paytoplay}.
By contrast, $x_{\text{lose}} = 0$, since there is no chance that ransomware operators will have to pay more than the cost to launch the attack -- the victim will simply ignore the attack because they do not value the information which has been ransomed or have some mitigation such as those outlined in Section~\ref{sec:lowering}.
In total, this means that the game played, from the perspective of ransomware operators, is as follows:
\begin{align*}
    L & = \{p_{\text{win}}, p_{\text{lose}}\} \\
    X & = \{x_{\text{win}}, 0\} \\
\end{align*}
and therefore, the expected utility for a single successful attack is:
\begin{align} 
 E[u(x)]    & = \sum_{i = \{\text{win}, \text{lose}\}} p_i (x_i - c) \nonumber \\
            & = (p_{\text{win}} (x_{\text{win}} - c)) + (p_{\text{lose}} (0 - c)) \nonumber \\
            & = p_{\text{win}} x_{\text{win}} - (p_{\text{win}} c + p_{\text{lose}} c) \nonumber \\
            & = p_{\text{win}} x_{\text{win}} - c \label{eqn:exp_util}
\end{align}

Since $x_{\text{lose}} = 0$ and $p_{\text{lose}} = 1 - p_{\text{win}}$, for the sake of simplicity and readability, we use $x$ and $p$ in the remainder of the paper to represent the case when a victim pays.
We can see from Equation~\ref{eqn:exp_util} that ransomware operators are incentivized to continue operating for as long as the value of $p x > c$, since they will profit from each attack, on average.
Research by Kaspersky Labs~\cite{kaspersky2021consumer} shows that 56\% of ransomware victims pay the ransom to restore access to their data. 
At this rate of payment, the cost of an average ransomware attack would need to be 1.7857 times -- nearly double -- the optimal payment to remove the incentive.

We can see that probabilistically, this is equivalent to betting on a biased coin flip.
Since $E[u(x)]$ is a function of the random variable $x$, it is itself a random variable, which we denote $Y$.
Given a cost to make a bet $c$, we flip a biased coin with win probability $p$ and receive payout $x$ at that rate.
Let $b$ be the amount of capital available to the bettor -- our attacker -- and let $b > c$.
We initialize $b_0$ to be the amount of capital available before any bets are cast and $b_i$ the available capital to the bettor at trial $i$.
Then after the first trial, our possible values for $b_1$ are $b_1 = b_0 - c$ or $b_1 = b_0 - c + x$. 
Our expected value of $b_1 = (b_0 - c) + p x$, as in Equation~\ref{eqn:exp_util}.

By the linearity of expectation, our expected bank at trial $k$ is:
\[b_k = b_0 + E[Y_k] = b_0 + k(p x - c)\]
We can see that if $px > c$, then the expected value of each trial is positive, and so for the player making the bet,
\begin{equation}
\lim_{k \rightarrow \infty} E[Y_k] = k(p x - c) = \infty \label{eqn:inf_exp}
\end{equation}
This suggests that any player who can participate in the game is highly incentivized to play as many rounds as possible, since the potential payoff is infinite.
Note that this expected value only holds in an idealized world with infinite money and no law enforcement, so it does not capture the intricate relationships of the real world.
It does, however, demonstrate that since the expectation is not finite, there is no optimal stopping time.
Therefore, there is no incentive for any attacker to ever stop conducting ransomware attacks when $px - c$ is reasonably large.

\begin{figure}
    \centering
    \includegraphics[width=\textwidth]{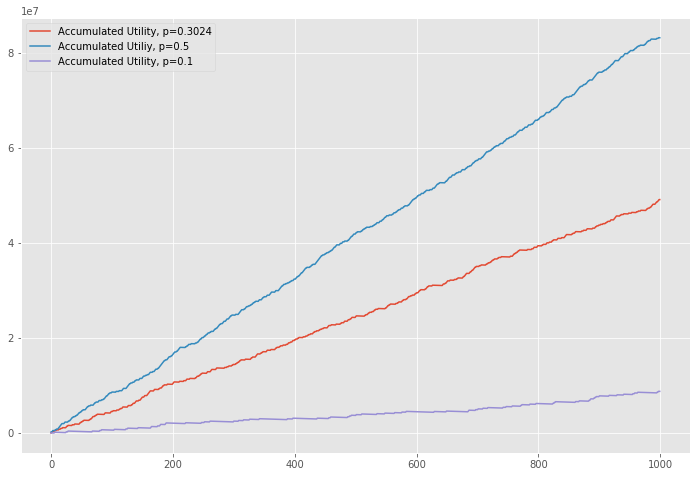}
    \caption{Plot of simulation demonstrating accumulated utility at p=0.1, p=0.3024, and p=0.5}
    \label{fig:accumutil}
\end{figure}

To demonstrate this, we construct three simple simulations, shown in Figure~\ref{fig:accumutil}.
We set our payout value $x=170404$ and cost $c=4200$ based on analysis in Section~\ref{sec:paytoplay}.
Then, for three different values of $p$: 0.1, 0.3024, and 0.5, we run 1000 trials.
With probability $p$, the player receives value $x - c$, and with probabiltiy $1-p$, the player receives value $-c$.
We can see that overall, the accumulated value is linear with respect to $p$, as we would expect from Equation~\ref{eqn:exp_util}.

\section{Paying to Play} \label{sec:paytoplay}
The cost of running a ransomware attack is very opaque and highly variable.
Some cybercriminal organizations are sophisticated operations that develop their malware in-house~\cite{doj2021high}.
These organizations have software development lifecycles, version control, testing, and pay staff to perform all of these functions.
Other organizations simply purchase ransomware-as-a-service~\cite{meland2020ransomware} (RaaS) or piece together their arsenal from so-called darknet markets.
A 2017 study~\cite{carbonblack2017dark} found that prices ranged from \$0.50 to \$3,000 for ransomware products, at a median price of \$10.50.
In contrast to these prices, most RaaS providers take a percentage of the ransom, rather than providing an executable for a flat fee.

In order to infect an endpoint with ransomware, however, one needs to gain initial access.
Furthermore, most ransomware operators leverage a loader -- a small program designed to install another malware on a target system -- to actually get the ransomware onto the endpoint. 
Nearly all ransomware variants~\cite{pan2021ransomware} rely on phishing, commodity malware, exploit kits, and vulnerable services -- particularly the remote desktop protocol -- to deliver their malware.
This factors in to the overall cost of operation, but is challenging to estimate, since cybercriminals are not forthcoming with this information.
A technical report issued by Deloitte~\cite{deloitte2018black} found the cost of initial access to be between \$70 and \$400 per 1000 machines depending on geographic region, and the cost of a loader to range from \$3 to \$4,000, depending on functionality.
The United States demanded the highest fee for an initial access at \$400.
At this time, the US is also the nation which demands the highest ransoms, and so in the interest of creating a conservative but accurate estimate, we use this number.
The highest average monthly cost of a loader was \$800, which is the figure we use moving forward.
We thus estimate the cost of an attack at $c = 3000 + 400 + 800 = 4200$.

This cost of \$4,200 means at at a payment rate of $p = 0.56$, the minimal ransom to turn a profit is \$7,500.
However, this payment rate is too large, since it assumes that the attack has been successful.
According to Sophos~\cite{sophos2021state}, only 54\% of attacks actually encrypt data.
Given that a successful attack is a precondition for being a paying victim, the joint probability of the attack being successful and the ransom being paid, which we defined in Equation~\ref{eqn:exp_util} as $p_{\text{win}}$ is the product of these two probabilities.
Our joint probability for a successful attack where the victim pays the ransom is therefore:
\[p = P(\text{paid} | \text{success}) \cdot P(\text{success}) = 0.56 \cdot 0.54 = 0.3024\]

This suggests that at a cost of \$4,200, per attack the minimal ransom an attacker must request to remain profitable is \$13,888.89.
As of March 2021, the average value of ransomware a payout for a compromised organization was \$312,493~\cite{cluley2021average}, around 22 times the minimal value needed to incentivize the attacks.
We note that other estimates, such as those by Sophos~\cite{sophos2021state} are a more modest \$170,404 for mid-sized organizations in the United states, a value which is still around 12 times the minimum to create positive expected value for these attacks.
We treat these as a ``reasonable average range'' in our subsequent analysis.

There are three variables in this problem that may disincentivize the perpetration of ransomware attacks:
\begin{enumerate}
    \item Lowering the value of the payments
    \item Increasing the cost of operating ransomware
    \item Decreasing the probability of payment
\end{enumerate}

We discuss the feasibility of using each of these three variables to disincentivize ransomware attacks in turn.

\subsection{Lowering the Value of Payments}
Today, there are few options for lowering the value of a payment.
Since nearly all payments for ransomware are rendered in cryptocurrency, a steep decline in the value of cryptocurrency or the inability to exchange it for other goods or services would remove the effective value of a successful attack. 
To date, some proposals have been made to ban~\cite{clark2021what}, or regulate cryptocurrencies~\cite{nabilou2019regulate,schaupp2018cryptocurrency}, though the effect of these bans and proposed regulations on the price of cryptocurrency remains to be seen.
Moreover, even if cryptocurrency were regulated into obsolescence, ransoms could be paid in gift cards or other hard to track currency equivalents.
This suggests that lowering the value of payments is not a viable path for removing the incentive.

\subsection{Increasing Costs}
The onus for increasing costs falls on the ransomware developers and operators themselves, and so there is likely a cost ceiling.
If the marketplace efficiencies of initial access brokers and ransomware-as-a-service were removed entirely, the cost of conducting an attack would be the cost of development plus the cost of deployment and maintenance of the infrastructure.
This would require more technical skill and initial investment than relatively low-skill ransomware operators would be capable of, but after the initial investment, would likely cost less per-attack than the \$3,000 high-end figure from~\cite{carbonblack2017dark}.
This may, on balance, reduce the overall prevalence of malware attacks.
However, this would also require the takedown of nearly all darknet marketplaces.
Despite a number of high-profile takedowns, ransomware continues to flourish on these marketplaces. 
Thus, the options for increasing costs to operators are also limited.

\subsection{Decreasing Payment Probability} \label{sec:payout}
Since the probability of payment is the one thing out of the control of the attackers, it stands to reason that it is where defenders can exercise the most control.
In our model, decreasing the probability of a successful attack that gets paid linearly reduces the expected value of an attack. 
This means that organizations have two options available to them to reduce an attack's expected value. 
Decreasing the success of launched attacks will prevent the victim having to decide whether or not to pay the ransom in the first place.
Assuming an attack is successful, decreasing the chance that the ransom is paid will also reduce the attacker's value.

Given our average payout value range of $x = [170,404, 312493]$, the expected value of an attack at current payment rates is in the range $[47,300.17, 170,798.08]$.
A 50\% reduction in probability of payout to $p = 0.28$ against a cost of $c = 4200$, with attack success rates held equal yields an expected value range of $[21565.08, 43048.94]$ -- an amount that a would-be ransomware operator could make as a software engineer in Europe~\cite{orosz2021trimodal} instead of perpetrating ransomware attacks.
Given the financial motivation of most ransomware operators~\cite{anderson2020security}, it stands to reason that a comparable salary is a perfectly substitutable good for rational actors.
To eliminate profit entirely, assuming current attack success rates and sufficient economies of scale, payment probability would need to decrease to 2.489\% on the high-end of average payments and 4.564\% on the low-end of payments -- a dramatic reduction from today's payment rates.

Despite that ``break-even'' probability, ransomware operators are likely to turn to some other income stream before profits hit zero due to law enforcement activities surrounding cybercrime.
In particular, the US Federal Bureau of Investigations and the UK National Cyber Security Centre have pursued cybercriminals abroad~\cite{bbc2021ransomware}, indicting and sanctioning ransomware operators.
However, in order to drastically reduce the payout rate of ransomware, organizations will need to have a reason not to pay the ransoms.

\section{Lowering the Stakes} \label{sec:lowering}
In order to lower the probability of payment and create an environment where attackers are not incentivized to continue launching ransomware attacks, victims must be incentivized not to pay the ransom.
An effective strategy for lowering the probability of payment ultimately consists of one where the victim's options for restoration are meaningfully less costly than paying the ransom. 
Considerable work has been done on quantifying these differences and we point to the article by Cluley~\cite{cluley2021average} for details, as the specific rates will differ from organization to organization.
Since the use of ransomware is illegal, there are external, non-financial mechanisms for reducing attacker incentives such as arrest, seizure of assets, indictment, and sanctions.
We do not address these mechanisms in our framework and reserve their impact for future work.

In order to reduce attacker incentives, we consider the potential impact of four commonly discussed strategies:
\begin{enumerate}
    \item Decreasing Attack Success
    \item Cyber Insurance
    \item Use of Decrypters
    \item Off-Site Backups
\end{enumerate}

\subsection{Decreasing Attack Success}
Decreasing attack success is the goal of any organizational information security program.
The success of attacks has myriad factors, ranging from human factors such as insider threats and phishing to software vulnerabilities and misconfigurations.
Modern antivirus technologies can assist in catching the loaders that often deliver the ransomware, and some endpoint security solutions can even detect exploitation of vulnerabilities. 
In addition, training programs for phishing emails and advising customers not to open attachments from unknown senders are widely used to attempt to mitigate these attacks.
A comprehensive listing of ways to reduce an organization's attack surface is out of the scope of this paper, but a 2020 report by Deloitte and the Financial Services Information Sharing and Analysis Center~\cite{bernard2020reshaping} showed that on average, 10\% of an organization's information technology budget -- approximately 0.2\% of company revenue -- is dedicated to cybersecurity.
In light of the increasing threats associated with ransomware, this amount may not be sufficient to reduce the probability that an attack is successful.

The figure in Equation~\ref{eqn:exp_util} only holds for cases where a ransomware infection has been successful and does not account for failed attacks -- only payments.
Reducing the incidence of these attacks through other means such as the use of application allowlists, strong spam filters, protection of exposed ports and services, and other well-known security hygiene methods can serve to reduce the success of these attacks. 
Since the cost to an attacker is undertaken whether or not the attack is successful, the failure of these attacks will discourage these attackers.
In order to isolate the influence of payment probability, our analysis assumed that all attacks are successful -- a naive assumption that suggests the 1.5\% payout probability derived in Section~\ref{sec:payout} is the probability of payment overall, not merely the conditional probability of payment given a successful attack.

\subsection{Cyber Insurance}
Cyber insurance is a strategy that is often mentioned as an organizational solution in the context of ransomware.
This can help to protect businesses from the cost of ransomware attacks, covering the cost to restore encrypted data.
However, in cases where cyber insurance alleviates the burden to victims, attackers are still paid, doing nothing to remove the incentives surrounding ransomware. 
Consequently, from an attacker incentive perspective, cyber insurance does nothing to alleviate the overall problem of ransomware.

\subsection{Use of Decrypters}
The use of decrypters is a significant way to allow victims to ignore the effects of ransomware. 
Although decrypters for some of the most popular strains of ransomware today are not available, organizations like No More Ransom!\footnote{https://www.nomoreransom.org} offers free decrypters for more than 150 families of ransomware.
Widespread knowledge of these utilities and increased investment by security researchers on developing these utilities could allow victims to decrypt their own files without paying a ransom.
Note that when decrypters become available or kill-switches as seen in WannaCry~\cite{mohurle2017brief} shut down operations, ransomware operators will patch their malware~\cite{arghire2017patched} to continue operations.

\subsection{Off-Site Backups}
The most commonly proposed solution for organizations to avoid the impacts of ransomware and confidently be able to not pay a ransom is the use of off-site backups. 
An off-site backup can be used to restore systems to pre-ransomware configurations and tends to cost significantly less than paying the ransom.
Research by Wood \textit{et al.}~\cite{wood2010disaster} acknowledges the difficulties of backup deployments.
Although they develop their recovery from a disaster preparedness perspective, their cost estimates show that both cloud-based and colocation for backups can allow for high uptime at a fraction of the cost associated with paying a ransom.
Additionally, having a backup that allows for restoration reduces the cost to remediate possible residual traces of the attacker, reduces time to remediate, and mitigates much of the reputational damage associated with paying a ransom.

\subsection{Impact of Mitigations}
The aforementioned approaches may allow victims to choose not to pay, but as Cartwright \textit{et al.}~\cite{cartwright2019pay} demonstrate, victims will have different willingness to pay given some set ransom.
This willingness to pay depends on the size of the ransom and therefore encourages the victim to mitigate the attack. 
When victims pay, they usually -- though not always~\cite{sophos2021state} -- get their files back, a factor which discourages paying.
However, there is some cost to deterrence, and if that is too high, the victim will instead accept their chances of being infected.

There are also factors at play external to the relationship between the cost of a ransom versus the cost of mitigation.
For example, in the United States, ransom payments can be written off~\cite{wood2020garmin} as ``ordinary, necessary, and reasonable'' expenses for tax purposes.
This factor actually incentivizes victims to pay, and discourages additional investments into mitigation.
Wheeler and Martin~\cite{wheeler2021should} point out that in the current regulatory environment of the United States, there is a misalignment between public interests to discourage ransomware and private interests to recover data and resume operations at the lowest cost.
We conclude then, that government and regulatory organizations interested in preventing ransomware should create financial incentives for organizations and individuals to invest in backups that allow for ransoms not to be paid.
Further, policy solutions to change the tax incentives associated with paying ransoms could be pursued to improve the chance that companies will invest in security technologies.

\section{Conclusion}
Ransomware remains a significant problem in the world, and our analysis demonstrates why -- there is effectively unlimited incentive to use ransomware.
Since the cost is relatively low and the potential payouts are high, financially-motivated actors are encouraged to pursue this line of attack.
Additionally, the victims of successful attacks are more likely to pay than not for a variety of factors, including the ability to write-off the ransom as a business expense. 

If we wish to eliminate the threat of ransomware, we cannot attack the market itself, as the actors are aware that their actions are illegal but have accepted that risk.
Instead, we must see that attackers are engaged in a simple game where they do not need to account for the strategies of their victims.
Where defenders have power to affect ransomware is largely on the front of actually paying the ransoms.

We outlined a handful of commonly-discussed solutions and conclude that off-site backups remain the most effective way to ignore the impact of ransomware attacks.
In order to encourage organizations to pursue these policies, we conclude that governmental and regulatory organizations will need to provide incentives for organizations to invest in these backup solutions.
Short of encouraging these solutions and allowing victims not to pay ransoms, we can reasonably expect the ransomware threat to continue to grow.

The model used here leveraged a probabilistic model and expected utility theory to identify incentives and explore the security impacts of those incentives.
In future work, we seek to explore a more realistic model of the risk behaviors these attackers and defenders exhibit based on their subjective beliefs.
Furthermore, there are meaningful non-financial mechanisms such as those mentioned in Section~\ref{sec:lowering}, and inclusion of those mechanisms would require a more complex model.
This could be done by representing uncertainty via cumulative prospect theory~\cite{tversky1992advances}, as has been done in the economic literature.
In particular, there is a significant amount of uncertainty on the part of attackers about whether or not an attack will be successful. 
Similarly, there is significant uncertainty for defenders about how, when, and where they will be attacked.
By representing the choice under uncertainty more richly than in an expected utility model, we may better model the true behaviors of attackers and defenders.

\bibliographystyle{splncs04}
\bibliography{references}
\end{document}